\documentclass[12pt,preprint,notoc,nohyper]{MCPlane1} 

\usepackage{epsfig,multicol,bbm}

\newcommand\fverb{\setbox\pippobox=\hbox\bgroup\verb}
\newcommand\fverbdo{\egroup\medskip\noindent%
            \fbox{\unhbox\pippobox}\ }
\newcommand\fverbit{\egroup\item[\fbox{\unhbox\pippobox}]}
\newbox\pippobox

\title{Lie symmetries of the energy-momentum tensor for plane symmetric static
spacetimes}

\author{K. Saifullah\\
    Department of Mathematics, Quaid-i-Azam University, Islamabad, Pakistan\\
    Electronic address: \email{saifullah@qau.edu.pk}}

\preprint{}  

\abstract{Matter collineations (MCs) are the vector fields along
which the energy-momentum tensor remains invariant under the Lie
transport. Invariance of the metric, the Ricci and the Riemann
tensors have been studied extensively and the vectors along which
these tensors remain invariant are called Killing vectors (KVs),
Ricci collineations (RCs) and curvature collineations (CCs),
respectively. In this paper plane symmetric static spacetimes have
been studied for their MCs. Explicit form of MCs together with the
Lie algebra admitted by them has been presented. Examples of
spacetimes have been constructed for which MCs have been compared
with their RCs and KVs. The comparison shows that neither of the
sets of RCs and MCs contains the other, in general.}

\dedicated{}

\begin{document}

\section{Introduction}

One of the principal applications of Lie derivatives in theoretical
physics is to express the notion that a tensor field is invariant
under some transformation. We say that a tensor field $\mathbf{T}$
is invariant under a vector field $\mathbf{V}$ if
\begin{equation}
\pounds _{\mathbf{V}}\mathbf{T}=0\,,
\end{equation}
where $\pounds _{\mathbf{V}}$ denotes Lie differentiation with
respect to the vector $\mathbf{V}$. If $\mathbf{T}$ has physical
importance then those special vector fields under which $\mathbf{T}$
is invariant will also be important. The manifolds of interest in
theoretical physics have metrics, and it is therefore of
considerable interest whenever the metric is invariant with respect
to some vector field. These vector fields are called the Killing
vectors (KVs) or isometries. After the spacetime metric, the
curvature, the Ricci and the energy-momentum tensors are other
important candidates which play a significant role in understanding
the geometric structure and physical properties of spacetimes in
relativity. While the isometries, provide information of the
symmetries inherent in the spacetime, the symmetries of the
matter-energy field are provided by MCs, vector fields along which
the energy momentum tensor is invariant under the Lie transport.
These symmetry properties are described by continuous groups of
motions or collineations and they lead to conservation laws. For an
introduction to spacetime symmetries and their significance the
reader may see References \cite{eisen, AQW, HS}.

Formally, the KV is defined as follows. A manifold $M$ is said to
admit a KV (or motion) $\xi ^{a}$ if the Lie derivative of the
metric $g_{ab}\;$with respect to $\xi ^{a}$ is conserved, i.e.

\begin{equation}
\pounds _{\mathbf{\xi }}g_{ab}=0\;.  \label{f14}
\end{equation}
The vector $\xi ^{a}$ is a Ricci collineation (RC) if the Lie
derivative of the Ricci tensor, $R_{ab}$, with respect to. $\xi
^{a}$ is conserved, i.e.
\begin{equation}
\pounds _{\mathbf{\xi}} R_{ab}=0\; .  \label{f17}
\end{equation}
Since the Ricci tensor is built from the metric tensor, it must
inherit its symmetries. Thus if the Lie derivative of $g_{ab}$
vanishes, it must vanish for $R_{ab}$ also. Hence every KV is an RC
but the converse may not be true. The RCs which are not KVs are
called proper RCs \cite{QSZ}. For plane symmetric spacetimes the RCs
are finite if the Ricci tensor is non-degenerate; for degenerate
case the RCs may be finite as well as
infinite dimensional \cite{QSZ, 15}. If $R_{ab}$ in Eq. $\left( \ref{f17}%
\right) $ is replaced by the energy-momentum tensor, $T_{ab}$, then
the vector \textbf{$\xi $ }is called an MC. In component form this
can be expressed as
\begin{equation}
\xi ^{c}T_{ab,c}+T_{ac}\xi _{,b}^{c}+T_{bc}\xi _{,a}^{c}=0\;.
\label{f18}
\end{equation}
Clearly, every KV is an MC also but the converse is not true.
Recently, MCs for different spacetimes have been discussed in the
literature \cite{HS, KS-th}. However, very little is known on this
important subject, particularly, the relationship between RCs and
MCs and there is a need for more research \cite{AQ}. In this paper
we study the Lie symmetries of the energy--momentum tensor, called
the matter collineations (MCs), for plane symmetric static
spacetimes, and compare them with their KVs and RCs \cite {QSZ, 15,
plnkv}.

The plan of the paper is as follows. In Sections 2 and 3 we
construct and solve the MC\ equations. Section 5 contains the
algebra of the MCs obtained in the previous section. Examples of the
metrics are given in Section 6, where MCs are compared with their
RCs and KVs. Concluding remarks are given at the end.

\section{The matter collineation equations}

We take $\left( x_{,}^{0}x_{,}^{1}x_{,}^{2}x^{3}\right) =\left(
t,x,y,z\right) $, so that, the most general plane symmetric static
line element can be written as \cite{mac}

\begin{equation}
ds^{2}=e^{\nu \left( x\right) }dt^{2}-dx^{2}-e^{\mu \left( x\right)
}\left( dy^{2}+dz^{2}\right) ,  \label{1}
\end{equation}
where $\nu $ and $\mu $ are functions of $x$ only. For this metric
the non-vanishing components of $T_{ab}$ are
\begin{equation}
\left.
\begin{array}{l}
T_{00}=-\frac{e^{\nu \left( x\right) }}{4}\left( 4\mu ^{\prime
\prime }+3\mu
^{\prime ^{2}}\right) , \\
T_{11}=\frac{1}{4}\left( 2\nu ^{\prime }\mu ^{\prime }+\mu ^{\prime
^{2}}\right), \\
T_{22}=\frac{e^{\mu \left( x\right) }}{4}\left( 2\nu ^{\prime \prime
}+2\mu ^{\prime \prime }+\nu ^{\prime ^{2}}+\mu ^{\prime ^{2}}+\nu
^{\prime }\mu ^{\prime }\right) =T_{33},
\end{array}
\right.  \label{stress}
\end{equation}
and those of the Ricci tensor are
\begin{equation}
\left.
\begin{array}{l}
R_{00}=\frac{e^{\nu }}{4}\left( 2\nu ^{\prime \prime }+\nu ^{\prime
^{2}}+2\nu ^{\prime }\mu ^{\prime }\right), \\
R_{11}=-\left( \frac{\nu ^{\prime \prime }}{2}+\mu ^{\prime \prime }+\frac{%
\nu ^{\prime ^{2}}}{4}+\frac{\mu ^{\prime ^{2}}}{2}\right) , \\
R_{22}=-\frac{e^{\mu }}{4}\left( 2\mu ^{\prime \prime }+\nu ^{\prime
}\mu ^{\prime }+2\mu ^{\prime ^{2}}\right) =R_{33} .
\end{array}
\right.  \label{2}
\end{equation}
Here $^{,\prime ,}$\ denotes differentiation with respect to $x$.
The Ricci scalar is given by
\begin{equation}
R=\nu ^{\prime \prime }+2\mu ^{\prime \prime }+\frac{1}{2}\left( \nu
^{\prime ^{2}}+3\mu ^{\prime ^{2}}+2\nu ^{\prime }\mu ^{\prime
}\right).
\end{equation}
Writing $T_{ii}=T_{i}$, for $i=0,1,2,3,$ the MC\ equations Eq.
$\left( \ref {f18}\right)$ for the MC vector {$\mathbf{\xi }$} take
the form

\begin{eqnarray}
T_{0}^{^{\prime }}\xi ^{1}+2T_{0}\xi _{,0}^{0} &=&0 ,  \label{3} \\
T_{0}\xi _{,1}^{0}+T_{1}\xi _{,0}^{1} &=&0,  \label{4} \\
T_{0}\xi _{,2}^{0}+T_{2}\xi _{,0}^{2} &=&0,  \label{5} \\
T_{0}\xi _{,3}^{0}+T_{2}\xi _{,0}^{3} &=&0,  \label{6} \\
T_{1}^{^{\prime }}\xi ^{1}+2T_{1}\xi _{,1}^{1} &=&0,  \label{7} \\
T_{1}\xi _{,2}^{1}+T_{2}\xi _{,1}^{2} &=&0,  \label{8} \\
T_{1}\xi _{,3}^{1}+T_{2}\xi _{,1}^{3} &=&0,  \label{9} \\
T_{2}^{^{\prime }}\xi ^{1}+2T_{2}\xi _{,2}^{2} &=&0,  \label{10} \\
T_{2}\left( \xi _{,3}^{2}+\xi _{,2}^{3}\right) &=&0,  \label{11} \\
T_{2}^{^{\prime }}\xi ^{1}+2T_{2}\xi _{,3}^{3} &=&0.  \label{12}
\end{eqnarray}

These are ten non-linear coupled partial differential equations
for$\,\xi
^{0}$, $\xi ^{1}$, $\xi ^{2}$, $\xi ^{3}$, and $T_{0}$, $T_{1}$, $T_{2}$, $%
T_{3}$. The $\xi ^{i}$ depend on $t$, $x$, $y$ and $z$; and the $T_{i}\,$on $%
x$ only.

\section{Solution of the matter collineation equations}

Solving Eqs. $\left( \ref{3}\right) $- $\left( \ref{12}\right) $
simultaneously we obtain the components of the MC vector,
\textbf{$\xi $}. Since the procedure for the solution of similar
systems of partial differential equations has appeared in the
literature \cite{QSZ, BQZ}, we do not give the calculations and only
present the results. The procedure, roughly speaking, is as follows.
We first integrate any of these equations to obtain the components
of \textbf{$\xi $}$\mathbf{\,}$in terms of
arbitrary functions of the coordinates. Using this form of \textbf{$\xi $}$%
\mathbf{\,}$in other MC equations will give conditions on these
arbitrary functions. Going back and forth in this way and checking
consistency with the MC equations at every step until these
functions are determined explicitly yields the final form of
\textbf{$\xi $} involving arbitrary constants. In the course of
finding these solutions we get constraints on the components of
$\mathbf{T}$. Thus we will arrive at various cases of MCs
corresponding to these constraints. Solving these constraints, which
are often differential in nature, gives the spacetimes. Now, we list
these cases.

\bigskip

{\Large Case 1 }

\smallskip

\textbf{Constraints on }$T_{ab}$\textbf{: }

\[
T_{0}^{\prime }\neq 0,\,\left( \frac{T_{2}^{\prime }}{T_{2}\sqrt{T_{1}}}%
\right) ^{\prime }\neq 0,\,\left( \frac{T_{0}}{T_{2}}\right)
^{\prime }=0.\,
\]

\textbf{MCs: }Let $\frac{T_{0}}{T_{2}}=k\,.$%
\begin{equation}
\left.
\begin{array}{l}
\xi ^{0}=-\frac{1}{k}\left( c_{6}z+c_{5}y\right) +c_{1}, \\
\xi ^{1}=0, \\
\xi ^{2}=c_{4}z+kc_{5}t+c_{2}, \\
\xi ^{3}=-yc_{4}+kc_{6}t+c_{3}.
\end{array}
\right.  \label{279}
\end{equation}

{\Large Case 2 }

\smallskip

\textbf{Constraints on }$T_{ab}$\textbf{: }

\[
T_{0}^{\prime }\neq 0,\,\left( \frac{T_{2}^{\prime }}{T_{2}\sqrt{T_{1}}}%
\right) ^{\prime }\neq 0,\left( \frac{T_{0}}{T_{2}}\right) ^{\prime
}\neq 0.\,
\]

\textbf{MCs: }
\begin{equation}
\left.
\begin{array}{l}
\xi ^{0}=c_{1}, \\
\xi ^{1}=0, \\
\xi ^{2}=c_{4}z+c_{2}, \\
\xi ^{3}=-c_{4}y+c_{3}.
\end{array}
\right.  \label{280}
\end{equation}

{\Large Case 3 }

\smallskip

\textbf{Constraints on }$T_{ab}$\textbf{: }
\[
T_{0}^{\prime }\neq 0,\left( \frac{T_{2}^{\prime }}{T_{2}\sqrt{T_{1}}}%
\right) ^{\prime }=0,\,T_{2}^{\prime }\neq 0,\,\left( \frac{T_{2}}{T_{0}}%
\right) ^{^{\prime }}\neq 0,\,\left( \frac{T_{0}^{\prime }}{T_{0}\sqrt{T_{1}}%
}\right) ^{\prime }\neq 0.\,
\]

\textbf{MCs: }
\begin{equation}
\left.
\begin{array}{l}
\xi ^{0}=c_{1}, \\
\xi ^{1}=0, \\
\xi ^{2}=c_{4}z+c_{2}, \\
\xi ^{3}=-c_{4}y+c_{3}.\thinspace
\end{array}
\right.  \label{301}
\end{equation}

{\Large Case 4 }

\smallskip

\textbf{Constraints on }$T_{ab}$\textbf{: }
\[
T_{0}^{\prime }\neq 0,\left( \frac{T_{2}^{\prime }}{T_{2}\sqrt{T_{1}}}%
\right) ^{\prime }=0,\,T_{2}^{\prime }\neq 0,\,\left( \frac{T_{2}}{T_{0}}%
\right) ^{^{\prime }}\neq 0,\left( \frac{T_{0}^{\prime }}{T_{0}\sqrt{T_{1}}}%
\right) ^{\prime }=0.
\]

\textbf{MCs: }Let$\frac{R_{2}^{\prime }}{R_{2}\sqrt{R_{1}}}=\alpha $, $\,%
\frac{T_{0}^{\prime }}{T_{0}\sqrt{T_{1}}}=\beta \neq \alpha \,$, where $%
\alpha $ and $\beta $ are constants.
\begin{equation}
\left.
\begin{array}{l}
\xi ^{0}=\frac{\beta }{\alpha }c_{5}t+c_{1}, \\
\xi ^{1}=-c_{5}\frac{2}{\alpha \sqrt{T_{1}}}, \\
\xi ^{2}=c_{4}z+c_{5}y+c_{2}\thinspace , \\
\xi ^{3}=-c_{4}y+c_{5}z+c_{3}.
\end{array}
\right.  \label{309}
\end{equation}

{\Large Case 5 }

\smallskip

\textbf{Constraints on }$T_{ab}$\textbf{: }
\[
T_{0}^{\prime }\neq 0,\left( \frac{T_{2}^{\prime }}{T_{2}\sqrt{T_{1}}}%
\right) ^{\prime }=0,\,T_{2}^{\prime }\neq 0,\left( \frac{T_{2}}{T_{0}}%
\right) ^{^{\prime }}=0.\,
\]

\textbf{MCs: }Here we write $\frac{R_{2}^{\prime }}{R_{2}\sqrt{R_{1}}}%
=\alpha $, $T_{2}=-\delta T_{0}$\thinspace .
\begin{equation}
\left.
\begin{array}{l}
\xi ^{0}=\frac{c_{7}}{2}\left( t^{2}-\frac{4}{\alpha
^{2}T_{0}}+\delta y^{2}+\delta z^{2}\right) +c_{6}\delta
z-c_{8}yt+c_{9}tz+c_{5}\delta
y+c_{10}t+c_{1}, \\
\xi ^{1}=-\frac{2}{\alpha \sqrt{T}_{1}}\left(
c_{7}t-c_{8}y+c_{9}z+c_{10}\right), \\
\xi ^{2}=c_{7}yt+\frac{c_{8}}{2}\left( -\frac{t^{2}}{\delta }+\frac{4}{%
\alpha ^{2}T_{2}}-y^{2}+z^{2}\right)
+c_{5}t+c_{9}yz+c_{10}y+c_{4}z+c_{2},
\\
\xi ^{3}=c_{7}zt-c_{8}yz-\frac{c_{9}}{2}\left( -\frac{t^{2}}{\delta }+\frac{4%
}{\alpha ^{2}T_{2}}+y^{2}-z^{2}\right) +c_{6}t+c_{10}z-c_{4}y+c_{3}.
\end{array}
\right.  \label{332}
\end{equation}

{\Large Case 6 }

\smallskip

\textbf{Constraints on }$T_{ab}$\textbf{: }
\[
T_{0}^{\prime }\neq 0,\,T_{2}^{\prime }=0,\left( \frac{\left( \sqrt{T_{0}}%
\right) ^{\prime }}{\sqrt{T_{1}}}\right) ^{\prime }=0\,.\,
\]

\textbf{MCs: }Put\textbf{\ }$\frac{\left( \sqrt{T_{0}}\right) ^{\prime }}{%
\sqrt{T_{1}}}=\gamma $ a constant.
\begin{equation}
\left.
\begin{array}{l}
\xi ^{0}=\frac{1}{\sqrt{T_{0}}}[ z\left( c_{7}\sin \gamma
t-c_{8}\cos \gamma t\right) +y\left( c_{5}\sin \gamma t-c_{6}\cos
\gamma t\right)\\
-\left( c_{9}\sin \gamma t-c_{10}\cos \gamma t\right)] +c_{1}, \\
\xi ^{1}=-\frac{1}{\sqrt{T_{1}}}[ z\left( c_{7}\cos \gamma
t+c_{8}\sin \gamma t\right) +y\left( c_{5}\cos \gamma t+c_{6}\sin
\gamma t\right)\\
-\left( c_{9}\cos \gamma t+c_{10}\sin \gamma t\right)], \\
\xi ^{2}=\frac{\sqrt{T_{0}}}{\gamma T_{2}}\left( c_{5}\cos \gamma
t+c_{6}\sin \gamma t\right) +c_{4}z+c_{2}, \\
\xi ^{3}=\frac{\sqrt{T_{0}}}{\gamma T_{2}}\left( c_{7}\cos \gamma
t+c_{8}\sin \gamma t\right) -c_{4}y+c_{3}.
\end{array}
\right.  \label{353}
\end{equation}

{\Large Case 7 }

\smallskip

\textbf{Constraints on }$T_{ab}$\textbf{: }
\[
T_{0}^{\prime }\neq 0,\,T_{2}^{\prime }=0,\left( \frac{\left( \sqrt{T_{0}}%
\right) ^{\prime }}{\sqrt{T_{1}}}\right) ^{\prime }\neq 0,\left[ \frac{T_{0}%
}{2\sqrt{T_{1}}}\left( \frac{T_{0}^{\prime
}}{T_{0}\sqrt{T_{1}}}\right) ^{\prime }\right] ^{\prime }=0.\,\,
\]

\textbf{MCs: }This implies that$\,\frac{T_{0}^{\prime }}{T_{0}\sqrt{T_{1}}}=$%
constant $=\lambda \neq 0\,$.
\begin{equation}
\left.
\begin{array}{l}
\xi ^{0}=c_{5}\left( \frac{1}{\lambda T_{0}}-\frac{\lambda
}{4}t^{2}\right)
-c_{6}\frac{\lambda }{2}t+c_{1}, \\
\xi ^{1}=\frac{1}{\sqrt{T_{1}}}\left( c_{5}t+c_{6}\right), \\
\xi ^{2}=c_{4}z+c_{2}, \\
\xi ^{3}=-c_{4}y+c_{3\,}.
\end{array}
\right.  \label{372}
\end{equation}

{\Large Case 8 }

\smallskip

\textbf{Constraints on }$T_{ab}$\textbf{: }
\[
T_{0}^{\prime }\neq 0,\,T_{2}^{\prime }=0,\left( \frac{\left( \sqrt{T_{0}}%
\right) ^{\prime }}{\sqrt{T_{1}}}\right) ^{\prime }\neq 0,\left[ \frac{T_{0}%
}{\sqrt{T_{1}}}\left( \frac{T_{0}^{\prime
}}{T_{0}\sqrt{T_{1}}}\right) ^{\prime }\right] ^{\prime }\neq 0.\,
\]

\textbf{MCs: }
\begin{equation}
\left.
\begin{array}{l}
\xi ^{0}=c_{1}, \\
\xi ^{1}=0, \\
\xi ^{2}=c_{4}z+c_{2}, \\
\xi ^{3}=-c_{4}y+c_{3}.
\end{array}
\right.  \label{382}
\end{equation}

{\Large Case 9 }

\smallskip

\textbf{Constraints on }$T_{ab}$\textbf{: }
\[
T_{0}^{\prime }=0,T_{2}^{\prime }=0.\,
\]

\textbf{MCs: }We put $T_{0}=-\alpha $ and $T_{2}=\beta $.
\begin{equation}
\left.
\begin{array}{l}
\xi ^{0}=-\frac{c_{6}}{\alpha }\int \sqrt{T_{1}}dx+c_{7}y+c_{8}z+c_{1}, \\
\xi ^{1}=\frac{1}{\sqrt{T_{1}}}\left( c_{6}t+c_{9}y+c_{10}z+c_{5}\right), \\
\xi ^{2}=\frac{\alpha }{\beta }c_{7}t-\frac{1}{\beta }c_{9}\int
\sqrt{T_{1}}
dx+c_{4}z+c_{2}, \\
\xi ^{3}=\frac{\alpha }{\beta }c_{8}t-\frac{1}{\beta }c_{10}\int
\sqrt{T_{1}} dx-\beta c_{4}y+c_{3}.
\end{array}
\right.  \label{B29}
\end{equation}

{\Large Case 10 }

\smallskip

\textbf{Constraints on }$T_{ab}$\textbf{: }
\[
T_{2}^{\prime }\neq 0,\,\left[ \frac{T_{2}}{\sqrt{T_{1}}}\left( \frac{%
T_{2}^{\prime }}{T_{2}\sqrt{T_{1}}}\right) ^{\prime }\right]
^{\prime }\neq 0.\,
\]

\textbf{MCs: }
\begin{equation}
\left.
\begin{array}{l}
\xi ^{0}=c_{1}, \\
\xi ^{1}=0, \\
\xi ^{2}=c_{4}z+c_{2}, \\
\xi ^{3}=-c_{4}y+c_{3}.
\end{array}
\right.  \label{B119}
\end{equation}

{\Large Case 11 }

\smallskip

\textbf{Constraints on }$T_{ab}$\textbf{: }
\[
T_{2}^{\prime }\neq 0,\,\left[ \frac{T_{2}}{\sqrt{T_{1}}}\left( \frac{%
T_{2}^{\prime }}{2T_{2}\sqrt{T_{1}}}\right) ^{\prime }\right]
^{\prime }=0,\,\left( \frac{T_{2}^{\prime
}}{T_{2}\sqrt{T_{1}}}\right) ^{\prime }\neq 0\,.\,
\]

\textbf{MCs: }
\begin{equation}
\left.
\begin{array}{l}
\xi ^{0}=c_{1}, \\
\xi ^{1}=0, \\
\xi ^{2}=c_{4}z+c_{2}, \\
\xi ^{3}=-c_{4}y+c_{3}.
\end{array}
\right.
\end{equation}

{\Large Case 12 }

\smallskip

\textbf{Constraints on }$T_{ab}$\textbf{: }
\[
T_{0}^{\prime }=0,T_{2}^{\prime }\neq 0,\,\left( \frac{T_{2}^{\prime }}{T_{2}%
\sqrt{T_{1}}}\right) ^{\prime }=0\,.\,
\]

\textbf{MCs: }Put $\frac{T_{2}^{\prime }}{T_{2}\sqrt{T_{1}}}=k_{1}$
\begin{equation}
\left.
\begin{array}{l}
\xi ^{0}=c_{1}, \\
\xi ^{1}=\frac{1}{\sqrt{T_{1}}}\left( c_{6}y+c_{7}z+c_{5}\right), \\
\xi ^{2}=-c_{6}\left( \int
\frac{\sqrt{T_{1}}}{T_{2}}dx+k_{1}\frac{y^{2}}{2}
-k_{1}\frac{z^{2}}{2}\right) -k_{1}c_{7}yz-k_{1}c_{5}y\;-c_{4}z+c_{2}, \\
\xi ^{3}=-k_{1}c_{6}yz-c_{7}\left( \int \frac{\sqrt{T_{1}}}{T_{2}}dx-\frac{%
k_{1}}{2}y^{2}+k_{1}\frac{z^{2}}{2}\right)
-k_{1}c_{5}z+c_{4}y\;+c_{3}.
\end{array}
\right.  \label{B206}
\end{equation}

For the degenerate $T_{ab}$, i.e. when det$\left( T_{ab}\right) =0$,
we get
MCs admitting infinite dimensional Lie algebras, except in one case when $%
T_{1}=0,\,T_{i}\neq 0$ for $i=0,2,3$. Cases 13 and 14 have
degenerate energy-momentum tensor but admit finite MCs.

\medskip

{\Large Case 13 }

\smallskip \textbf{Constraints on }$T_{ab}$\textbf{: }

\[
T_{1}=0,\,T_{0}\neq 0,T_{2}\neq 0,\,\left( \frac{T_{0}^{^{\prime }}T_{2}}{%
T_{0}T_{2}^{\prime }}\right) ^{^{\prime }}=0\,.\,
\]

\textbf{MCs: Put }$\frac{T_{0}^{^{\prime }}T_{2}}{T_{0}T_{2}^{\prime }}%
=k_{2} $.
\begin{equation}
\left.
\begin{array}{l}
\xi ^{0}=c_{5}t+c_{1}, \\
\xi ^{1}=-c_{5}\frac{2T_{0}}{T_{0}^{\prime }}, \\
\xi ^{2}=c_{4}z+\frac{1}{k_{2}}c_{5}y+c_{2}, \\
\xi ^{3}=-c_{4}y+\frac{1}{k_{2}}c_{5}z+c_{3}.
\end{array}
\right.
\end{equation}

{\Large Case 14 }

\smallskip

\textbf{Constraints on }$T_{ab}$\textbf{: }
\[
T_{1}=0,\,T_{0}\neq 0,T_{2}\neq 0,T_{0}^{^{\prime }}\neq
0,T_{2}^{^{\prime }}\neq 0,\,\left( \frac{T_{0}}{T_{2}}\right)
^{^{\prime }}=0\,.\,
\]

\textbf{MCs: }Here we write $T_{0}=-\alpha T_{2}$\thinspace .
\begin{equation}
\left.
\begin{array}{l}
\xi ^{0}=c_{7}\left( \frac{y^{2}}{2}+\frac{z^{2}}{2}+\alpha
\frac{t^{2}}{2}
\right) +c_{8}ty+c_{5}y+c_{9}tz+c_{6}z+c_{10}t+c_{1}, \\
\xi ^{1}=-\frac{2T_{0}}{T_{0}^{\prime }}\left( \alpha
c_{7}t+c_{8}y+c_{9}z+c_{10}\right), \\
\xi ^{2}=\alpha c_{7}ty+c_{8}\left( \alpha
\frac{t^{2}}{2}+\frac{y^{2}}{2}-
\frac{z^{2}}{2}\right) +\alpha c_{5}t+c_{9}yz+c_{10}y-c_{4}z+c_{2}, \\
\xi ^{3}=\alpha c_{7}tz+c_{8}yz+c_{9}\left( \alpha \frac{t^{2}}{2}-\frac{%
y^{2}}{2}+\frac{z^{2}}{2}\right) +\alpha
c_{6}t+c_{10}z+c_{4}y+c_{3}.
\end{array}
\right.  \label{2.41}
\end{equation}

{\Large Case 15 }

\smallskip

If either $T_{0}$ or $T_{2}$ (or both) are zero $T_{ab}$ becomes
degenerate and MCs admit infinite dimensional Lie algebra.

\section{Lie algebras of matter collineations}

If a set of vector fields on a manifold under the operation of Lie
bracket (defined by the Lie derivative on a manifold) satisfies the
conditions of anti-commutativity and Jacobi's identity, one gets a
Lie algebra. Here we provide the Lie algebraic structure for the MC\
vector fields obtained in the last section and identify their
nature. We also classify them into solvable and semisimple algebras
and identify some of their sub-algebras.

\bigskip

{\Large Case 1 }

\smallskip

Generators:\

$\hspace{0.75in}\left.
\begin{array}{l}
\mathbf{X}_{1}=\partial _{t}\,, \\
\mathbf{X}_{2}=\partial _{y}\,, \\
\mathbf{X}_{3}=\partial _{z}\,, \\
\mathbf{X}_{4}=z\partial _{y}\,-y\partial _{z}\,, \\
\mathbf{X}_{5}=y\partial _{t}+k^{2}t\partial _{y}\,, \\
\mathbf{X}_{6}=z\partial _{t}+k^{2}t\partial _{z}\,.
\end{array}
\right. $

Algebra:
\[
\begin{tabular}{lll}
$\left[ \mathbf{X}_{1},\mathbf{X}_{5}\right] =k^{2}\mathbf{X}_{2}\,,$ & $%
\left[ \mathbf{X}_{1},\mathbf{X}_{6}\right] =k^{2}\mathbf{X}_{3}\,,$ & $%
\left[ \mathbf{X}_{2},\mathbf{X}_{4}\right] =-\mathbf{X}_{3}\,,$ \\
$\left[ \mathbf{X}_{2},\mathbf{X}_{5}\right] =\mathbf{X}_{1}\,,$ &
$\left[
\mathbf{X}_{3},\mathbf{X}_{4}\right] =\mathbf{X}_{2}\,,$ & $\left[ \mathbf{X}%
_{3},\mathbf{X}_{6}\right] =\mathbf{X}_{1}\,,$ \\
$\left[ \mathbf{X}_{4},\mathbf{X}_{5}\right] =\mathbf{X}_{6}\,,$ &
$\left[
\mathbf{X}_{4},\mathbf{X}_{6}\right] =-\mathbf{X}_{5}\,,$ & $\left[ \mathbf{X%
}_{5},\mathbf{X}_{6}\right] =-k^{2}\mathbf{X}_{4}\,,$ \\
$\left[ \mathbf{X}_{i},\mathbf{X}_{j}\right] =0\,,\,
\mathrm{otherwise.} $ & &
\end{tabular}
\]

This is $SO(1,2)\times \left[ SO(2)\otimes \Bbb{R}^{2}\right] $ where `$%
\times $' represents the semi-direct and `$\otimes $' the direct
product. Here $\mathbf{X}_{5}$ and $\mathbf{X}_{6}$ are the Lorentz
boosts in $y$ and $z$ directions. $\mathbf{X}_{4}$ is a rotation in
$y$ and $z$. This is a
semisimple algebra having $\left\langle \mathbf{X}_{4},\mathbf{X}_{5},%
\mathbf{X}_{6}\right\rangle $ as a subalgebra. \bigskip

{\Large Case 2 }

\smallskip

Generators:\

$\hspace{0.75in}\left.
\begin{array}{l}
\mathbf{X}_{1}=\partial _{t}\,, \\
\mathbf{X}_{2}=\partial _{y}\,, \\
\mathbf{X}_{3}=\partial _{z}\,, \\
\mathbf{X}_{4}=z\partial _{y}-y\partial _{z}\,.
\end{array}
\right. $

Algebra:
\[
\begin{tabular}{lll}
$\left[ \mathbf{X}_{2},\mathbf{X}_{4}\right] =-\mathbf{X}_{3}\,,$ &
$\left[
\mathbf{X}_{3},\mathbf{X}_{4}\right] =\mathbf{X}_{2}\,,$ & $\left[ \mathbf{X}%
_{i},\mathbf{X}_{j}\right] =0\,,\,\mathrm{otherwise. }$%
\end{tabular}
\]

This can be written as $\left\{ SO(2)\times \left[ \Bbb{R}\otimes
SO(2)\right] \right\} \otimes \Bbb{R}$ and is solvable. \bigskip

{\Large Case 3}

\smallskip

Generators:\

$\hspace{0.75in}\left.
\begin{array}{l}
\mathbf{X}_{1}=\partial _{t}\,, \\
\mathbf{X}_{2}=\partial _{y}\,, \\
\mathbf{X}_{3}=\partial _{z}\,, \\
\mathbf{X}_{4}=z\partial _{y}-y\partial _{z}\,.
\end{array}
\right. $

Algebra:
\[
\begin{tabular}{lll}
$\left[ \mathbf{X}_{2},\mathbf{X}_{4}\right] =-\mathbf{X}_{3}\,,$ &
$\left[
\mathbf{X}_{3},\mathbf{X}_{4}\right] =\mathbf{X}_{2}\,,$ & $\left[ \mathbf{X}%
_{i},\mathbf{X}_{j}\right] =0\,,\,\mathrm{otherwise. }$%
\end{tabular}
\]

This is the same as the previous case. \bigskip

{\Large Case 4 }

\smallskip

Generators:\

$\hspace{0.75in}\left.
\begin{array}{l}
\mathbf{X}_{1}=\partial _{t}\,, \\
\mathbf{X}_{2}=\partial _{y}\,, \\
\mathbf{X}_{3}=\partial _{z}\,, \\
\mathbf{X}_{4}=z\partial _{y}-y\partial _{z}\,, \\
\mathbf{X}_{5}=\frac{\beta }{\alpha }t\partial _{t}-\frac{2}{\alpha \sqrt{%
T_{1}}}\partial _{x}+y\partial _{y}+z\partial _{z}\,.
\end{array}
\right. $

Algebra:
\[
\begin{tabular}{lll}
$\left[ \mathbf{X}_{1},\mathbf{X}_{5}\right] =\frac{\beta }{\alpha }\mathbf{X%
}_{1}\,,$ & $\left[ \mathbf{X}_{2},\mathbf{X}_{4}\right]
=-\mathbf{X}_{3}\,,$
& $\left[ \mathbf{X}_{2},\mathbf{X}_{5}\right] =\mathbf{X}_{2}\,,$ \\
$\left[ \mathbf{X}_{3},\mathbf{X}_{4}\right] =\mathbf{X}_{2}\,,$ &
$\left[
\mathbf{X}_{3},\mathbf{X}_{5}\right] =\mathbf{X}_{3}\,,$ & $\left[ \mathbf{X}%
_{i},\mathbf{X}_{j}\right] =0\,,\,\mathrm{otherwise. }$%
\end{tabular}
\]

This is a solvable algebra which can be written as $G_{5}=\left\langle G_{4},%
\mathbf{X}_{5}\right\rangle $, where

$G_{4}=\left\{ SO(2)\times \left[ \Bbb{R}\otimes SO(2)\right]
\right\} \otimes \Bbb{R}\mathsf{.\bigskip }$

{\Large Case 5 }

\smallskip

Generators:\

$\hspace{0.75in}\left.
\begin{array}{l}
\mathbf{X}_{1}=\partial _{t}\,, \\
\mathbf{X}_{2}=\partial _{y}\,, \\
\mathbf{X}_{3}=\partial _{z}\,, \\
\mathbf{X}_{4}=z\partial _{y}-y\partial _{z}\,\,, \\
\mathbf{X}_{5}=\delta y\partial _{t}+t\partial _{y}\,, \\
\mathbf{X}_{6}=\delta z\partial _{t}+t\partial _{z}\,, \\
\mathbf{X}_{7}=\frac{1}{2}\left( t^{2}-\frac{4}{\alpha
^{2}T_{0}}+\delta
y^{2}+\delta z^{2}\right) \partial _{t}-\frac{2}{\alpha \sqrt{T_{1}}}%
t\partial _{x}+yt\partial _{y}+zt\partial _{z}\,, \\
\mathbf{X}_{8}=-yt\partial _{t}+\frac{2}{\alpha \sqrt{T_{1}}}y\partial _{x}+%
\frac{1}{2}\left( -\frac{t^{2}}{\delta }+\frac{4}{\alpha ^{2}T_{2}}%
-y^{2}+z^{2}\right) \partial _{y}-yz\partial _{z}\,, \\
\mathbf{X}_{9}=tz\partial _{t}-\frac{2}{\alpha
\sqrt{T_{1}}}z\partial
_{x}+yz\partial _{y}-\frac{1}{2}\left( -\frac{t^{2}}{\delta }+\frac{4}{%
\alpha ^{2}T_{2}}+y^{2}-z^{2}\right) \partial _{z}\,, \\
\mathbf{X}_{10}=t\partial _{t}-\frac{2}{\alpha \sqrt{T_{1}}}\partial
_{x}+y\partial _{y}+z\partial _{z}\,.
\end{array}
\right. $

Algebra:
\[
\begin{tabular}{lll}
$\left[ \mathbf{X}_{1},\mathbf{X}_{5}\right] =\mathbf{X}_{2}\,,$ &
$\left[
\mathbf{X}_{1},\mathbf{X}_{6}\right] =\mathbf{X}_{3}\,,$ & $\left[ \mathbf{X}%
_{1},\mathbf{X}_{7}\right] =\mathbf{X}_{10}\,,$ \\
$\left[ \mathbf{X}_{1},\mathbf{X}_{8}\right] =\frac{1}{\delta }\mathbf{X}%
_{5}\,,$ & $\left[ \mathbf{X}_{1},\mathbf{X}_{9}\right] =\frac{1}{\delta }%
\mathbf{X}_{6}\,,$ & $\left[ \mathbf{X}_{1},\mathbf{X}_{10}\right] =\mathbf{X%
}_{1}\,,$ \\
$\left[ \mathbf{X}_{2},\mathbf{X}_{4}\right] =-\mathbf{X}_{3},$ &
$\left[ \mathbf{X}_{2},\mathbf{X}_{5}\right] =\delta
\mathbf{X}_{1}\,,$ & $\left[
\mathbf{X}_{2},\mathbf{X}_{7}\right] =\mathbf{X}_{5}\,,$ \\
$\left[ \mathbf{X}_{2},\mathbf{X}_{8}\right] =-\mathbf{X}_{10}$ &
$\left[
\mathbf{X}_{2},\mathbf{X}_{9}\right] =\mathbf{X}_{4}\,,$ & $\left[ \mathbf{X}%
_{2},\mathbf{X}_{10}\right] =\mathbf{X}_{2}\,,$ \\
$\left[ \mathbf{X}_{3},\mathbf{X}_{4}\right] =\mathbf{X}_{2}\,,$ &
$\left[ \mathbf{X}_{3},\mathbf{X}_{6}\right] =\delta
\mathbf{X}_{1}\,,$ & $\left[
\mathbf{X}_{3},\mathbf{X}_{7}\right] =\mathbf{X}_{6}\,,$ \\
$\left[ \mathbf{X}_{3},\mathbf{X}_{8}\right] =\mathbf{X}_{4}\,,$ &
$\left[
\mathbf{X}_{3},\mathbf{X}_{9}\right] =\mathbf{X}_{10}\,,$ & $\left[ \mathbf{X%
}_{3},\mathbf{X}_{10}\right] =\mathbf{X}_{3}\,,$ \\
$\left[ \mathbf{X}_{4},\mathbf{X}_{5}\right] =\mathbf{X}_{6}\,,$ &
$\left[
\mathbf{X}_{4},\mathbf{X}_{6}\right] =-\mathbf{X}_{5}\,,$ & $\left[ \mathbf{X%
}_{4},\mathbf{X}_{8}\right] =-\mathbf{X}_{9}\,,$ \\
$\left[ \mathbf{X}_{4},\mathbf{X}_{9}\right] =\mathbf{X}_{8}\,,$ &
$\left[ \mathbf{X}_{5},\mathbf{X}_{6}\right] =-\delta
\mathbf{X}_{4}\,,$ & $\left[
\mathbf{X}_{5},\mathbf{X}_{7}\right] =-\delta \mathbf{X}_{8}\,,$ \\
$\left[ \mathbf{X}_{5},\mathbf{X}_{8}\right] =-\mathbf{X}_{7}\,,$ &
$\left[ \mathbf{X}_{6},\mathbf{X}_{7}\right] =\delta
\mathbf{X}_{9}\,,$ & $\left[
\mathbf{X}_{6},\mathbf{X}_{9}\right] =\mathbf{X}_{7}\,,$ \\
$\left[ \mathbf{X}_{7},\mathbf{X}_{10}\right] =-\mathbf{X}_{7}\,,$ &
$\left[
\mathbf{X}_{8},\mathbf{X}_{10}\right] =-\mathbf{X}_{8}\,,$ & $\left[ \mathbf{%
X}_{9},\mathbf{X}_{10}\right] =-\mathbf{X}_{9}\,,$ \\
$\left[ \mathbf{X}_{i},\mathbf{X}_{j}\right]
=0\,,\,\mathrm{otherwise. }$ & &
\end{tabular}
\]

This $SO(1,4)$ or $SO(2,3)$ is the maximal semisimple anti-de Sitter
algebra. It has 3 dimensional subalgebras $\left\{ \mathbf{X}_{4},\mathbf{X}%
_{5},\mathbf{X}_{6}\right\} $ of rotations and $\left\{ \mathbf{X}_{8},%
\mathbf{X}_{9},\mathbf{X}_{10}\right\} $; 4 dimensional subalgebras
$\left\{
\mathbf{X}_{1},\mathbf{X}_{2},\mathbf{X}_{3},\mathbf{X}_{4}\right\} $ and $%
\left\{
\mathbf{X}_{7},\mathbf{X}_{8},\mathbf{X}_{9},\mathbf{X}_{10}\right\}
$; and 6 dimensional subalgebras $\left\{ \mathbf{X}_{1},\mathbf{X}_{2},%
\mathbf{X}_{3},\mathbf{X}_{4},\mathbf{X}_{5},\mathbf{X}_{6}\right\}
$ in it.
\bigskip

\pagebreak

{\Large Case 6 }

\smallskip

Generators:\

$\hspace{0.75in}\left.
\begin{array}{l}
\mathbf{X}_{1}=\partial _{t}\,, \\
\mathbf{X}_{2}=\partial _{y}\,, \\
\mathbf{X}_{3}=\partial _{z}\,, \\
\mathbf{X}_{4}=z\partial _{y}-y\partial _{z}\,\,, \\
\mathbf{X}_{5}=\frac{1}{\sqrt{T_{0}}}y\sin \gamma t\partial _{t}-\frac{1}{%
\sqrt{T_{1}}}y\cos \gamma t\partial _{x}+\frac{\sqrt{T_{0}}}{\gamma T_{2}}%
\cos \gamma t\partial _{y}\,, \\
\mathbf{X}_{6}=-\frac{1}{\sqrt{T_{0}}}y\cos \gamma t\partial _{t}-\frac{1}{%
\sqrt{T_{1}}}y\sin \gamma t\partial _{x}+\frac{\sqrt{T_{0}}}{\gamma T_{2}}%
\sin \gamma t\partial _{y}\,, \\
\mathbf{X}_{7}=\frac{1}{\sqrt{T_{0}}}z\sin \gamma t\partial _{t}-\frac{1}{%
\sqrt{T_{1}}}z\cos \gamma t\partial _{x}+\frac{\sqrt{T_{0}}}{\gamma T_{2}}%
\cos \gamma t\partial _{z}\,, \\
\mathbf{X}_{8}=-\frac{1}{\sqrt{T_{0}}}z\cos \gamma t\partial _{t}-\frac{1}{%
\sqrt{T_{1}}}z\sin \gamma t\partial _{x}+\frac{\sqrt{T_{0}}}{\gamma T_{2}}%
\sin \gamma t\partial _{z}\,, \\
\mathbf{X}_{9}=-\frac{1}{\sqrt{T_{0}}}\sin \gamma t\partial _{t}+\frac{1}{%
\sqrt{T_{1}}}\cos \gamma t\partial _{x}\,, \\
\mathbf{X}_{10}=\frac{1}{\sqrt{T_{0}}}\cos \gamma t\partial _{t}+\frac{1}{%
\sqrt{T_{1}}}\sin \gamma t\partial _{x}\,.
\end{array}
\right. $

Algebra:

\[
\begin{tabular}{lll}
$\left[ \mathbf{X}_{1},\mathbf{X}_{5}\right] =-\mathbf{X}_{6}\,,$ &
$\left[
\mathbf{X}_{1},\mathbf{X}_{6}\right] =\mathbf{X}_{5}\,,$ & $\left[ \mathbf{X}%
_{1},\mathbf{X}_{7}\right] =-\mathbf{X}_{8}\,,$ \\
$\left[ \mathbf{X}_{1},\mathbf{X}_{8}\right] =\mathbf{X}_{7}\,,$ &
$\left[
\mathbf{X}_{1},\mathbf{X}_{9}\right] =-\mathbf{X}_{10}\,,$ & $\left[ \mathbf{%
X}_{1},\mathbf{X}_{10}\right] =\mathbf{X}_{9}\,,$ \\
$\left[ \mathbf{X}_{2},\mathbf{X}_{4}\right] =-\mathbf{X}_{3}\,,$ &
$\left[
\mathbf{X}_{2},\mathbf{X}_{5}\right] =-\mathbf{X}_{9}\,,$ & $\left[ \mathbf{X%
}_{2},\mathbf{X}_{6}\right] =-\mathbf{X}_{10}\,,$ \\
$\left[ \mathbf{X}_{3},\mathbf{X}_{4}\right] =\mathbf{X}_{2}\,,$ &
$\left[
\mathbf{X}_{3},\mathbf{X}_{7}\right] =-\mathbf{X}_{9}\,,$ & $\left[ \mathbf{X%
}_{3},\mathbf{X}_{8}\right] =-\mathbf{X}_{10}\,,$ \\
$\left[ \mathbf{X}_{4},\mathbf{X}_{5}\right] =\mathbf{X}_{7}\,,$ &
$\left[
\mathbf{X}_{4},\mathbf{X}_{6}\right] =\mathbf{X}_{8}\,,$ & $\left[ \mathbf{X}%
_{4},\mathbf{X}_{7}\right] =-\mathbf{X}_{5}\,,$ \\
$\left[ \mathbf{X}_{4},\mathbf{X}_{8}\right] =-\mathbf{X}_{6}\,,$ &
$\left[
\mathbf{X}_{5},\mathbf{X}_{6}\right] =-\frac{1}{\gamma T_{2}}\mathbf{X}%
_{1}\,,$ & $\left[ \mathbf{X}_{5},\mathbf{X}_{7}\right] =\frac{1}{T_{2}}%
\mathbf{X}_{4}\,,$ \\
$\left[ \mathbf{X}_{5},\mathbf{X}_{9}\right] =-\frac{1}{T_{2}}\mathbf{X}%
_{2}\,,$ & $\left[ \mathbf{X}_{6},\mathbf{X}_{8}\right] =\frac{1}{T_{2}}%
\mathbf{X}_{4}\,,$ & $\left[ \mathbf{X}_{6},\mathbf{X}_{10}\right] =-\frac{1%
}{T_{2}}\mathbf{X}_{2}\,,$ \\
$\left[ \mathbf{X}_{7},\mathbf{X}_{8}\right] =-\frac{1}{\gamma T_{2}}\mathbf{%
X}_{1}\,,$ & $\left[ \mathbf{X}_{7},\mathbf{X}_{9}\right] =-\frac{1}{T_{2}}%
\mathbf{X}_{3}\,,$ & $\left[ \mathbf{X}_{8},\mathbf{X}_{10}\right] =-\frac{1%
}{T_{2}}\mathbf{X}_{3}\,,$ \\
$\left[ \mathbf{X}_{i},\mathbf{X}_{j}\right]
=0\,,\,\mathrm{otherwise. }$ & &
\end{tabular}
\]

This is again a 10 dimensional semisimple algebra and has $\left\{ \mathbf{X}%
_{1},\mathbf{X}_{2},\mathbf{X}_{3},\mathbf{X}_{4}\right\} $ as a
subalgebra.
\bigskip

{\Large Case 7 }

\smallskip

Generators:\

$\hspace{0.75in}\left.
\begin{array}{l}
\mathbf{X}_{1}=\partial _{t}\,, \\
\mathbf{X}_{2}=\partial _{y}\,, \\
\mathbf{X}_{3}=\partial _{z}\,, \\
\mathbf{X}_{4}=z\partial _{y}-y\partial _{z}\,\,, \\
\mathbf{X}_{5}=\left( \frac{1}{\lambda T_{0}}-\frac{\lambda
}{4}t^{2}\right)
\partial _{t}+\frac{1}{\sqrt{T_{1}}}t\partial _{x}\,, \\
\mathbf{X}_{6}=-\frac{\lambda }{2}t\partial _{t}+\frac{1}{\sqrt{T_{1}}}%
\partial _{x}\,.
\end{array}
\right. $

\pagebreak

Algebra:

\[
\begin{tabular}{lll}
$\left[ \mathbf{X}_{1},\mathbf{X}_{5}\right] =\mathbf{X}_{6}\,,$ &
$\left[ \mathbf{X}_{1},\mathbf{X}_{6}\right] =-\frac{\lambda
}{2}\mathbf{X}_{1}\,,$
& $\left[ \mathbf{X}_{2},\mathbf{X}_{4}\right] =-\mathbf{X}_{3}\,,$ \\
$\left[ \mathbf{X}_{3},\mathbf{X}_{4}\right] =\mathbf{X}_{2}\,,$ &
$\left[ \mathbf{X}_{5},\mathbf{X}_{6}\right] =\frac{\lambda
}{2}\mathbf{X}_{5}\,,$ &
$\left[ \mathbf{X}_{i},\mathbf{X}_{j}\right] =0\,,\,\mathrm{otherwise. }$%
\end{tabular}
\]

This is a semisimple algebra having $\left\{ \mathbf{X}_{1},\mathbf{X}_{2},%
\mathbf{X}_{3},\mathbf{X}_{4}\right\} $ and $\left\{ \mathbf{X}_{1},\mathbf{X%
}_{5},\mathbf{X}_{6}\right\} $ as subalgebras. \bigskip

{\Large Case 8 }

\smallskip

Generators:\

$\hspace{0.75in}\left.
\begin{array}{l}
\mathbf{X}_{1}=\partial _{t}\,, \\
\mathbf{X}_{2}=\partial _{y}\,, \\
\mathbf{X}_{3}=\partial _{z}\,, \\
\mathbf{X}_{4}=z\partial _{y}-y\partial _{z}\,\,.
\end{array}
\right. $

Algebra:
\[
\begin{tabular}{lll}
$\left[ \mathbf{X}_{2},\mathbf{X}_{4}\right] =-\mathbf{X}_{3}\,,$ &
$\left[
\mathbf{X}_{3},\mathbf{X}_{4}\right] =\mathbf{X}_{2}\,,$ & $\left[ \mathbf{X}%
_{i},\mathbf{X}_{j}\right] =0\,,\,\mathrm{otherwise. }$%
\end{tabular}
\]

This is the same as in Case 2. \bigskip

{\Large Case 9 }

\smallskip

Generators:\

$\hspace{0.75in}\left.
\begin{array}{l}
\mathbf{X}_{1}=\partial _{t}\,, \\
\mathbf{X}_{2}=\partial _{y}\,, \\
\mathbf{X}_{3}=\partial _{z}\,, \\
\mathbf{X}_{4}=z\partial _{y}-\frac{\beta }{\gamma }y\partial _{z}\,\,, \\
\mathbf{X}_{5}=\frac{1}{\sqrt{T_{1}}}\partial _{x}\,, \\
\mathbf{X}_{6}=-\frac{1}{\alpha }\int \sqrt{T_{1}}dx\partial _{t}+\frac{1}{%
\sqrt{T_{1}}}t\partial _{x}\,, \\
\mathbf{X}_{7}=y\partial _{t}+\frac{\alpha }{\beta }t\partial _{y}\,, \\
\mathbf{X}_{8}=z\partial _{t}+\frac{\alpha }{\gamma }t\partial _{z}\,, \\
\mathbf{X}_{9}=\frac{y}{\sqrt{T_{1}}}\partial _{x}-\frac{1}{\beta
}\int
\sqrt{T_{1}}dx\partial _{y}\,, \\
\mathbf{X}_{10}=\frac{z}{\sqrt{T_{1}}}\partial _{x}-\frac{1}{\gamma
}\int \sqrt{T_{1}}dx\partial _{z}\,.
\end{array}
\right. $

\pagebreak

Algebra:

\[
\begin{tabular}{lll}
$\left[ \mathbf{X}_{1},\mathbf{X}_{6}\right] =\mathbf{X}_{5}\,,$ &
$\left[
\mathbf{X}_{1},\mathbf{X}_{7}\right] =\frac{\alpha }{\beta }\mathbf{X}%
_{2}\,, $ & $\left[ \mathbf{X}_{1},\mathbf{X}_{8}\right] =\frac{\alpha }{%
\gamma }\mathbf{X}_{3}\,,$ \\
$\left[ \mathbf{X}_{2},\mathbf{X}_{4}\right] =-\frac{\beta }{\gamma }\mathbf{%
X}_{3}\,,$ & $\left[ \mathbf{X}_{2},\mathbf{X}_{7}\right]
=\mathbf{X}_{1}\,,$
& $\left[ \mathbf{X}_{2},\mathbf{X}_{9}\right] =\mathbf{X}_{5}\,,$ \\
$\left[ \mathbf{X}_{3},\mathbf{X}_{4}\right] =\mathbf{X}_{2}\,,$ &
$\left[
\mathbf{X}_{3},\mathbf{X}_{8}\right] =\mathbf{X}_{1}\,,$ & $\left[ \mathbf{X}%
_{3},\mathbf{X}_{10}\right] =\mathbf{X}_{5}\,,$ \\
$\left[ \mathbf{X}_{4},\mathbf{X}_{7}\right] =\mathbf{X}_{8}\,,$ &
$\left[
\mathbf{X}_{4},\mathbf{X}_{8}\right] =-\frac{\beta }{\gamma }\mathbf{X}%
_{7}\,,$ & $\left[ \mathbf{X}_{4},\mathbf{X}_{9}\right]
=\mathbf{X}_{10}\,,$
\\
$\left[ \mathbf{X}_{4},\mathbf{X}_{10}\right] =-\frac{\beta }{\gamma }%
\mathbf{X}_{9}\,,$ & $\left[ \mathbf{X}_{5},\mathbf{X}_{6}\right] =\frac{1}{%
\alpha }\mathbf{X}_{1}\,,$ & $\left[ \mathbf{X}_{5},\mathbf{X}_{9}\right] =-%
\frac{1}{\beta }\mathbf{X}_{2}\,,$ \\
$\left[ \mathbf{X}_{5},\mathbf{X}_{10}\right] =-\frac{1}{\gamma }\mathbf{X}%
_{3}\,,$ & $\left[ \mathbf{X}_{6},\mathbf{X}_{7}\right]
=-\mathbf{X}_{9}\,,$
& $\left[ \mathbf{X}_{6},\mathbf{X}_{8}\right] =-\mathbf{X}_{10}\,,$ \\
$\left[ \mathbf{X}_{6},\mathbf{X}_{9}\right] =-\frac{1}{\alpha }\mathbf{X}%
_{7}\,,$ & $\left[ \mathbf{X}_{6},\mathbf{X}_{10}\right] =-\frac{1}{\alpha }%
\mathbf{X}_{8}\,,$ & $\left[ \mathbf{X}_{7},\mathbf{X}_{8}\right] =-\frac{%
\alpha }{\beta }\mathbf{X}_{4}\,,$ \\
$\left[ \mathbf{X}_{7},\mathbf{X}_{9}\right] =\frac{\alpha }{\beta }\mathbf{X%
}_{6}\,,$ & $\left[ \mathbf{X}_{8},\mathbf{X}_{10}\right] =\frac{\alpha }{%
\gamma }\mathbf{X}_{6}\,,$ & $\left[ \mathbf{X}_{9},\mathbf{X}_{10}\right] =%
\frac{1}{\beta }\mathbf{X}_{4}\,,$ \\
$\left[ \mathbf{X}_{i},\mathbf{X}_{j}\right]
=0\,,\,\mathrm{otherwise. }$ & &
\end{tabular}
\]

This semisimple algebra has a 3 dimensional subalgebra $\left\{ \mathbf{X}%
_{4},\mathbf{X}_{9},\mathbf{X}_{10}\right\} $ of rotations, 4
dimensional
subalgebra $\left\{ \mathbf{X}_{1},\mathbf{X}_{2},\mathbf{X}_{3},\mathbf{X}%
_{4}\right\} $ and 7 dimensional subalgebra $\left\{ \mathbf{X}_{1},\mathbf{X%
}_{2},\mathbf{X}_{3},\mathbf{X}_{4},\mathbf{X}_{5},\mathbf{X}_{9},\mathbf{X}%
_{10}\right\} $ in it. \bigskip

{\Large Case 10 }

\smallskip

Generators:\

$\hspace{0.75in}\left.
\begin{array}{l}
\mathbf{X}_{1}=\partial _{t}\,, \\
\mathbf{X}_{2}=\partial _{y}\,, \\
\mathbf{X}_{3}=\partial _{z}\,, \\
\mathbf{X}_{4}=z\partial _{y}-ky\partial _{z}\,\,.
\end{array}
\right. $

Algebra:
\[
\begin{tabular}{lll}
$\left[ \mathbf{X}_{2},\mathbf{X}_{4}\right] =-k\mathbf{X}_{3}\,,$ &
$\left[
\mathbf{X}_{3},\mathbf{X}_{4}\right] =\mathbf{X}_{2}\,,$ & $\left[ \mathbf{X}%
_{i},\mathbf{X}_{j}\right] =0\,,\,\mathrm{otherwise. }$%
\end{tabular}
\]
\bigskip

{\Large Case 11 }

\smallskip

Generators:\

$\hspace{0.75in}\left.
\begin{array}{l}
\mathbf{X}_{1}=\partial _{t}\,, \\
\mathbf{X}_{2}=\partial _{y}\,, \\
\mathbf{X}_{3}=\partial _{z}\,, \\
\mathbf{X}_{4}=z\partial _{y}-y\partial _{z}\,\,.
\end{array}
\right. $

Algebra:
\[
\begin{tabular}{lll}
$\left[ \mathbf{X}_{2},\mathbf{X}_{4}\right] =-\mathbf{X}_{3}\,,$ &
$\left[
\mathbf{X}_{3},\mathbf{X}_{4}\right] =\mathbf{X}_{2}\,,$ & $\left[ \mathbf{X}%
_{i},\mathbf{X}_{j}\right] =0\,,\,\mathrm{otherwise. }$%
\end{tabular}
\]

Its structure is similar to that of Case 2. \bigskip

{\Large Case 12 }

\smallskip

Generators:\

$\hspace{0.75in}\left.
\begin{array}{l}
\mathbf{X}_{1}=\partial _{t}\,, \\
\mathbf{X}_{2}=\partial _{y}\,, \\
\mathbf{X}_{3}=\partial _{z}\,, \\
\mathbf{X}_{4}=-z\partial _{y}+y\partial _{z}\,\,, \\
\mathbf{X}_{5}=\frac{1}{\sqrt{T_{1}}}\partial _{x}\,-k_{1}y\partial
_{y}-k_{1}z\partial _{z}\,, \\
\mathbf{X}_{6}=\frac{y}{\sqrt{T_{1}}}\partial _{x}-\left( \int \frac{\sqrt{%
T_{1}}}{T_{2}}dx+k_{1}\frac{y^{2}}{2}-k_{1}\frac{z^{2}}{2}\right)
\partial
_{y}-k_{1}yz\partial _{z}\,,\,\,\, \\
\mathbf{X}_{7}=\frac{z}{\sqrt{T_{1}}}\partial _{x}-k_{1}yz\partial
_{y}\;-\left( \int \frac{\sqrt{T_{1}}}{T_{2}}dx-\frac{k_{1}}{2d}y^{2}+k_{1}%
\frac{z^{2}}{2}\right) \partial _{z}\,.
\end{array}
\right. $

Algebra:

\[
\begin{tabular}{lll}
$\left[ \mathbf{X}_{2},\mathbf{X}_{4}\right] =\mathbf{X}_{3}\,,$ &
$\left[ \mathbf{X}_{2},\mathbf{X}_{5}\right]
=-k_{1}\mathbf{X}_{2}\,,$ & $\left[
\mathbf{X}_{2},\mathbf{X}_{6}\right] =\mathbf{X}_{5}\,,$ \\
$\left[ \mathbf{X}_{2},\mathbf{X}_{7}\right] =k_{1}\mathbf{X}_{4}\,,$ & $%
\left[ \mathbf{X}_{3},\mathbf{X}_{4}\right] =-\mathbf{X}_{2}\,,$ &
$\left[
\mathbf{X}_{3},\mathbf{X}_{5}\right] =-k_{1}\mathbf{X}_{3}\,,$ \\
$\left[ \mathbf{X}_{3},\mathbf{X}_{6}\right] =-k_{1}\mathbf{X}_{4}\,,$ & $%
\left[ \mathbf{X}_{3},\mathbf{X}_{7}\right] =\mathbf{X}_{5}\,,$ &
$\left[
\mathbf{X}_{4},\mathbf{X}_{6}\right] =-\mathbf{X}_{7}\,,$ \\
$\left[ \mathbf{X}_{4},\mathbf{X}_{7}\right] =-\mathbf{X}_{6}\,,$ &
$\left[
\mathbf{X}_{5},\mathbf{X}_{6}\right] =-\mathbf{X}_{6}\,,$ & $\left[ \mathbf{X%
}_{5},\mathbf{X}_{7}\right] =-\mathbf{X}_{7}\,,$ \\
$\left[ \mathbf{X}_{6},\mathbf{X}_{7}\right] =K\mathbf{X}_{4}\,,$ &
$\left[ \mathbf{X}_{i},\mathbf{X}_{j}\right]
=0\,,\,\mathrm{otherwise. }$ &
\end{tabular}
\]

where $K=-2k_{1}\left( \int \frac{\sqrt{T_{1}}}{T_{2}}dx+\frac{1}{2k_{1}T_{2}%
}\right) $ is a constant. This is a semisimple algebra having
$\left\{
\mathbf{X}_{1},\mathbf{X}_{2},\mathbf{X}_{3},\mathbf{X}_{4}\right\} $ and $%
\left\{
\mathbf{X}_{4},\mathbf{X}_{5},\mathbf{X}_{6},\mathbf{X}_{7}\right\}
$
as 4 dimensional subalgebras and a 6 dimensional subalgebra $\left\{ \mathbf{%
X}_{2},\mathbf{X}_{3},\mathbf{X}_{4},\mathbf{X}_{5},\mathbf{X}_{6},\mathbf{X}%
_{7}\right\} $ in it. We write this as $G_{7}=\left\langle G_{4},\mathbf{X}%
_{5},\mathbf{X}_{6},\mathbf{X}_{7}\right\rangle $ where
$G_{4}=\left\{
SO(2)\times \left[ \Bbb{R}\otimes SO(2)\right] \right\} \otimes \Bbb{R}%
\mathsf{.}$

\smallskip

\smallskip {\Large Case 13 }

\smallskip Generators:\

$\hspace{0.75in}\left.
\begin{array}{l}
\mathbf{X}_{1}=\partial _{t}\,, \\
\mathbf{X}_{2}=\partial _{y}\,, \\
\mathbf{X}_{3}=\partial _{z}\,, \\
\mathbf{X}_{4}=z\partial _{y}-y\partial _{z}, \\
\mathbf{X}_{5}=t\partial _{t}-\frac{2T_{0}}{T_{0}^{\prime }}\partial _{x}\,+%
\frac{1}{k_{2}}y\partial _{y}+\frac{1}{k_{2}}z\partial _{z}.
\end{array}
\right. $

Algebra:

\[
\begin{tabular}{lll}
$\left[ \mathbf{X}_{1},\mathbf{X}_{5}\right] =\mathbf{X}_{1}\,,$ &
$\left[
\mathbf{X}_{2},\mathbf{X}_{4}\right] =-\mathbf{X}_{3}\,,$ & $\left[ \mathbf{X%
}_{2},\mathbf{X}_{5}\right] =\frac{1}{k_{2}}\mathbf{X}_{2}\,,$ \\
$\left[ \mathbf{X}_{3},\mathbf{X}_{4}\right] =\mathbf{X}_{2}\,,$ &
$\left[
\mathbf{X}_{3},\mathbf{X}_{5}\right] =\frac{1}{k_{2}}\mathbf{X}_{3}\,,$ & $%
\left[ \mathbf{X}_{i},\mathbf{X}_{j}\right] =0\,,\,\mathrm{otherwise. }$%
\end{tabular}
\]

This is a solvable algebra which can be written as $G=\left\langle G_{4},%
\mathbf{X}_{5}\right\rangle $, where

$G_{4}=\left\{ SO(2)\times \left[ \Bbb{R}\otimes SO(2)\right]
\right\} \otimes \Bbb{R}\mathsf{.}$

\bigskip

\pagebreak

{\Large Case 14 }

\smallskip

\smallskip Generators:\

$\hspace{0.75in}\left.
\begin{array}{l}
\mathbf{X}_{1}=\partial _{t}\,, \\
\mathbf{X}_{2}=\partial _{y}\,, \\
\mathbf{X}_{3}=\partial _{z}\,, \\
\mathbf{X}_{4}=-z\partial _{y}+y\partial _{z}\,\,, \\
\mathbf{X}_{5}=y\partial _{t}+\alpha t\partial _{y}\,, \\
\mathbf{X}_{6}=z\partial _{t}+\alpha t\partial _{z}\,, \\
\mathbf{X}_{7}=\left( \frac{y^{2}}{2}+\frac{z^{2}}{2}+\alpha \frac{t^{2}}{2}%
\right) \partial _{t}-\frac{2\alpha T_{0}}{T_{0}^{\prime }}t\partial
_{x}+\alpha ty\partial _{y}+\alpha tz\partial _{z}\,\,, \\
\mathbf{X}_{8}=ty\partial _{t}-\frac{2T_{0}}{T_{0}^{\prime
}}y\partial _{x}+\left( \alpha
\frac{t^{2}}{2}+\frac{y^{2}}{2}-\frac{z^{2}}{2}\right)
\partial _{y}+yz\,\partial _{z}, \\
\mathbf{X}_{9}=tz\partial _{t}\,-\frac{2T_{0}}{T_{0}^{\prime
}}z\partial
_{x}+yz\partial _{y}+\left( \alpha \frac{t^{2}}{2}-\frac{y^{2}}{2}+\frac{%
z^{2}}{2}\right) \partial _{z}\,, \\
\mathbf{X}_{10}=t\partial _{t}-\frac{2T_{0}}{T_{0}^{\prime
}}\partial _{x}\,+y\partial _{y}+z\partial _{z}.
\end{array}
\right. $

Algebra:

\[
\begin{tabular}{lll}
$\left[ \mathbf{X}_{1},\mathbf{X}_{5}\right] =\alpha \mathbf{X}_{2}\,,$ & $%
\left[ \mathbf{X}_{1},\mathbf{X}_{6}\right] =\alpha \mathbf{X}_{3}\,,$ & $%
\left[ \mathbf{X}_{1},\mathbf{X}_{7}\right] =\alpha \mathbf{X}_{10}\,,$ \\
$\left[ \mathbf{X}_{1},\mathbf{X}_{8}\right] =\mathbf{X}_{5}\,,$ &
$\left[
\mathbf{X}_{1},\mathbf{X}_{9}\right] =\mathbf{X}_{6}\,,$ & $\left[ \mathbf{X}%
_{1},\mathbf{X}_{10}\right] =\mathbf{X}_{1}\,,$ \\
$\left[ \mathbf{X}_{2},\mathbf{X}_{4}\right] =\mathbf{X}_{3}\,,$ &
$\left[
\mathbf{X}_{2},\mathbf{X}_{5}\right] =\mathbf{X}_{1}\,,$ & $\left[ \mathbf{X}%
_{2},\mathbf{X}_{7}\right] =\mathbf{X}_{5}\,,$ \\
$\left[ \mathbf{X}_{2},\mathbf{X}_{8}\right] =\mathbf{X}_{10}\,,$ &
$\left[
\mathbf{X}_{2},\mathbf{X}_{9}\right] =-\mathbf{X}_{4}\,,$ & $\left[ \mathbf{X%
}_{2},\mathbf{X}_{10}\right] =\mathbf{X}_{2}\,,$ \\
$\left[ \mathbf{X}_{3},\mathbf{X}_{4}\right] =-\mathbf{X}_{2}\,,$ &
$\left[
\mathbf{X}_{3},\mathbf{X}_{6}\right] =\mathbf{X}_{1}\,,$ & $\left[ \mathbf{X}%
_{3},\mathbf{X}_{7}\right] =\mathbf{X}_{6}\,,$ \\
$\left[ \mathbf{X}_{3},\mathbf{X}_{8}\right] =\mathbf{X}_{4}\,,$ &
$\left[
\mathbf{X}_{3},\mathbf{X}_{9}\right] =\mathbf{X}_{10}\,,$ & $\left[ \mathbf{X%
}_{3},\mathbf{X}_{10}\right] =\mathbf{X}_{3}\,,$ \\
$\left[ \mathbf{X}_{4},\mathbf{X}_{5}\right] =-\mathbf{X}_{6}\,,$ &
$\left[
\mathbf{X}_{4},\mathbf{X}_{6}\right] =\mathbf{X}_{5}\,,$ & $\left[ \mathbf{X}%
_{4},\mathbf{X}_{8}\right] =-\mathbf{X}_{9}\,,$ \\
$\left[ \mathbf{X}_{4},\mathbf{X}_{9}\right] =\mathbf{X}_{8}\,,$ &
$\left[ \mathbf{X}_{5},\mathbf{X}_{6}\right] =\alpha
\mathbf{X}_{4}\,,$ & $\left[
\mathbf{X}_{5},\mathbf{X}_{7}\right] =\alpha \mathbf{X}_{8}\,,$ \\
$\left[ \mathbf{X}_{5},\mathbf{X}_{8}\right] =\mathbf{X}_{7}\,,$ &
$\left[
\mathbf{X}_{6},\mathbf{X}_{7}\right] =-\mathbf{X}_{9}\,,$ & $\left[ \mathbf{X%
}_{6},\mathbf{X}_{9}\right] =\mathbf{X}_{7}\,,$ \\
$\left[ \mathbf{X}_{7},\mathbf{X}_{10}\right] =-\mathbf{X}_{7}\,,$ &
$\left[
\mathbf{X}_{8},\mathbf{X}_{10}\right] =-\mathbf{X}_{8}\,,$ & $\left[ \mathbf{%
X}_{9},\mathbf{X}_{10}\right] =-\mathbf{X}_{9}\,,$ \\
$\left[ \mathbf{X}_{i},\mathbf{X}_{j}\right]
=0\,,\,\mathrm{otherwise. }$ & &
\end{tabular}
\]

It has 3 dimensional subalgebras $\left\{ \mathbf{X}_{4},\mathbf{X}_{5},%
\mathbf{X}_{6}\right\} $ of rotations and $\left\{ \mathbf{X}_{8},\mathbf{X}%
_{9},\mathbf{X}_{10}\right\} $; 4 dimensional subalgebras $\left\{ \mathbf{X}%
_{1},\mathbf{X}_{2},\mathbf{X}_{3},\mathbf{X}_{4}\right\} $ and
$\left\{
\mathbf{X}_{7},\mathbf{X}_{8},\mathbf{X}_{9},\mathbf{X}_{10}\right\}
$; and
6 dimensional subalgebras $\left\{ \mathbf{X}_{1},\mathbf{X}_{2},\mathbf{X}%
_{3},\mathbf{X}_{4},\mathbf{X}_{5},\mathbf{X}_{6}\right\} $ in it. This $%
SO(1,4)$ or $SO(2,3)$ anti-de Sitter Lie algebra is the maximal
semisimple algebra of the degenerate case.

\section{Examples of metrics}

Here we give a few examples of spacetimes comparing MCs with their
RCs \cite {QSZ, 15} and KVs \cite{plnkv}.

\begin{enumerate}
\item
\[
ds^{2}=e^{\nu }\left( dt^{2}-dy^{2}-dz^{2}\right) -dx^{2},(\nu
^{\prime \prime }\neq 0)\,.
\]

For this metric the non-vanishing components of $T_{ab}$ are
\begin{equation}
\left.
\begin{array}{l}
T_{00}=-\frac{e^{\nu }}{4}\left( 4\nu ^{\prime \prime }+3\nu
^{\prime
^{2}}\right) \,, \\
T_{11}=\frac{1}{4}\left( 3\mu ^{\prime ^{2}}\right) \,, \\
T_{22}=\frac{e^{\nu }}{4}\left( 4\nu ^{\prime \prime }+3\nu ^{\prime
^{2}}\right) =T_{33}\,,
\end{array}
\right.
\end{equation}
and those of the Ricci tensor are
\begin{equation}
\left.
\begin{array}{l}
R_{00}=\frac{e^{\nu }}{4}\left( 2\nu ^{\prime \prime }+3\nu ^{\prime
^{2}}\right) \,, \\
R_{11}=-3\left( \frac{\nu ^{\prime \prime }}{2}+\frac{\nu ^{\prime ^{2}}}{4}%
\right) \,, \\
R_{22}=-\frac{e^{\nu }}{4}\left( 2\mu ^{\prime \prime }+3\nu
^{\prime ^{2}}\right) =R_{33}\,.
\end{array}
\right.
\end{equation}

It has 6 KVs, 6 RCs and 6 MCs given in Case 1.

\item
\[
ds^{2}=e^{\nu }dt^{2}-dx^{2}-e^{\mu }\left( dy^{2}+dz^{2}\right) \,,
(\nu ^{\prime \prime }\neq 0, \mu ^{\prime \prime }\neq 0)\,.
\]

It admits 4 KVs, 4 RCs and 4 MCs given by Case 2.

\item
\[
ds^{2}=e^{^{Ax}}\left( dt^{2}-dy^{2}-dz^{2}\,\right) -dx^{2}\,.
\]

This is an anti-de Sitter metric admitting 10 KVs, 10 RCs and 10 MCs
given by Case 5.

\item
\[
ds^{2}=x^{a}dt^{2}-dx^{2}-x^{b}\left( dy^{2}+dz^{2}\right) \,.
\]
The $T_{ab}$ in this case are
\begin{equation}
\left.
\begin{array}{l}
T_{00}=\left( a-\frac{3}{4}a^{2}\right) x^{c-2}\,, \\
T_{11}=\left( \frac{ca}{2}+\frac{a^{2}}{4}\right) /x^{2}\,, \\
T_{22}=\frac{1}{4}\left( a^{2}+c^{2}-2c-2a+ac\right)
x^{a-2}=T_{33}\,.
\end{array}
\right.
\end{equation}

This is an example of Case 4 with 4 KVs, 5 RCs and 5 MCs.

\item
\[
ds^{2}=\left( x/x_{0}\right) ^{2a}dt^{2}-dx^{2}-\left(
x/x_{0}\right) ^{2}\left( dy^{2}+dz^{2}\,\right) \,,
\]

$a$ and $x_{0}$ are constants and $a\neq 0,1,-1$. For this metric
$R_{ab}$ are given by
\begin{equation}
\left.
\begin{array}{l}
R_{00}=a\left( 1+a\right) x^{2a-2}/x_{0}^{2a}\,, \\
R_{11}=-\left( -a+a^{2}\right) /x^{2}\,, \\
R_{22}=-\left( 1+a\right) /x_{0}^{2}=R_{33}.
\end{array}
\right.
\end{equation}
The energy-momentum tensor for this metric can be written as with
\begin{equation}
\left.
\begin{array}{l}
T_{00}=-x^{2a-2}/x_{0}^{2a}\,, \\
T_{11}=\left( 2a+1\right) /x^{2}\,, \\
T_{22}=a^{2}/x_{0}^{2}=T_{33}\,.
\end{array}
\right.
\end{equation}

It has 4 KVs, 6 RCs and 6 MCs (Case 7). The energy density is
negative and cannot be made positive by introducing a cosmological
constant, therefore, it is unphysical.

\item
\[
ds^{2}=dt^{2}-dx^{2}-dy^{2}-dz^{2}\,.
\]

For this metric $T_{ab}=R_{ab}=0$, for all $a$, $b$, which means
that every direction is an MC and similarly RCs are also arbitrary
functions of the coordinates $t$, $x$, $y$ and $z$ giving an
infinite dimensional Lie algebra. However, it has 10 KVs. This is a
wrapped Minkowski spacetime.

\item
\[
ds^{2}=dt^{2}-dx^{2}-e^{Ax}\left( dy^{2}-dz^{2}\right),
\]

$A$ is a non-zero constant. For this metric the non-vanishing components of $%
T_{ab}$ are
\begin{equation}
\left.
\begin{array}{l}
T_{00}=-\frac{3A^{2}}{4}, \\
T_{11}=\frac{A^{2}}{4}, \\
T_{22}=\frac{A^{2}e^{Ax}}{4}=T_{33},
\end{array}
\right.
\end{equation}
and those of the Ricci tensor are
\begin{equation}
\left.
\begin{array}{l}
R_{00}=0\,\,, \\
R_{11}=-\frac{A^{2}}{2}\,, \\
R_{22}=-\frac{A^{2}e^{Ax}}{2}=R_{33}\,.
\end{array}
\right.
\end{equation}
It admits 7 KVs, infinitely many RCs and 7 MCs, given in Case 12.

\item
\[
ds^{2}=e^{Ax}dt^{2}-dx^{2}-e^{Bx}\left( dy^{2}+dz^{2}\right) \,,
\]

For this metric the non-vanishing components of $T_{ab}$ are
\begin{equation}
\left.
\begin{array}{l}
T_{00}=-\frac{3B^{2}e^{Ax}}{4}\,, \\
T_{11}=\frac{1}{4}\left( 2AB+B^{2}\right) \,\,, \\
T_{22}=\frac{e^{Bx}}{4}\left( A^{2}+B^{2}+AB\right) =T_{33}\,,
\end{array}
\right.
\end{equation}
and those of the Ricci tensor are
\begin{equation}
\left.
\begin{array}{l}
R_{00}=\frac{e^{Ax}}{4}\left( A^{2}+2AB\right) \,, \\
R_{11}=-\left( \frac{A^{2}}{4}+\frac{B^{2}}{2}\right) \,, \\
R_{22}=-\frac{e^{Bx}}{4}\left( AB+2B^{2}\right) =R_{33}\,.
\end{array}
\right.
\end{equation}
Now, if $A\neq B$ it admits 5 KVs, 5 RCs and 5 MCs, otherwise it has
10 RCs and 10 MCs.

\item
\[
ds^{2}=\left( x/x_{0}\right) ^{2a}dt^{2}-dx^{2}-\left(
x/x_{0}\right) ^{4/3}\left( dy^{2}+dz^{2}\,\right) \,,
\]

$a$ and $x_{0}$ are constants and $a\neq 0,1,-1$. The
energy-momentum tensor for this metric can be written as with
\begin{equation}
\left.
\begin{array}{l}
T_{00}=0\,, \\
T_{11}=16/9x^{2}\,, \\
T_{22}=-\left( a^{2}-\frac{a}{3}-\frac{2}{9}\right)
/x^{2/3}x_{0}^{4/3}=T_{33}.
\end{array}
\right.
\end{equation}
For this metric $R_{ab}$ are given by
\begin{equation}
\left.
\begin{array}{l}
R_{00}=a\left( a+\frac{1}{3}\right) x^{2a-2}/x_{0}^{2a}\,, \\
R_{11}=\left( a-a^{2}-\frac{4}{3}\right) /x^{2}\,, \\
R_{22}=-\frac{2}{3}\left( a+3\right) /x^{2/3}x_{0}^{4/3}=R_{33}.
\end{array}
\right.
\end{equation}

This space admits 4 KVs, 5 RCs and infinitely many MCs (Case 15).
\end{enumerate}

\section{Conclusion}

The plane symmetric static spacetimes have been studied for their
MCs. The MC equations have been solved giving rise to various cases
characterized by the constraints on the components of the
energy-momentum tensor. This includes cases of the non-degenerate as
well the degenerate tensor. Their Lie algebra structure has also
been given. Particular examples of metrics have been provided for
which MCs have been compared with their KVs and RCs. In some spaces
the RCs are greater than the MCs, while in others the MCs are more
than the RCs, which shows that neither of the sets contains the
other, in general.

\acknowledgments

The author is grateful to Asghar Qadir for reading the manuscript
and suggesting some improvements. He would also like to thank M.
Ziad, U. Camci and M. Sharif for useful discussions.

\end{document}